\documentclass[prl,aps,longbibliography,floats,superscriptaddress,twocolumn]{revtex4-1}

\usepackage[utf8]{inputenc}
\usepackage{amsmath,amssymb,amsthm,amsbsy,bm}
\usepackage{dsfont,verbatim,varwidth,dcolumn}
\usepackage{graphicx,epsfig,epstopdf,subfigure}
\usepackage[usenames,dvipsnames]{xcolor}
\usepackage{subfiles}

\def\ua{{\uparrow}}
\def\da{{\downarrow}}
\newcommand{\phm}{{\phi_{m}}}
\newcommand{\vcr}{\vec{r}}

\makeatother

\begin{document}

\title{Fractional edge reconstruction in integer quantum Hall phases}
\author{Udit Khanna}
\affiliation{Raymond and Beverly Sackler School of Physics and Astronomy, Tel Aviv University, Tel Aviv 6997801, Israel}
\affiliation{Department of Condensed Matter Physics, Weizmann Institute of Science, Rehovot 76100, Israel}
\author{Moshe Goldstein}
\affiliation{Raymond and Beverly Sackler School of Physics and Astronomy, Tel Aviv University, Tel Aviv 6997801, Israel}
\author{Yuval Gefen}
\affiliation{Department of Condensed Matter Physics, Weizmann Institute of Science, Rehovot 76100, Israel}

\date{\today}

\begin{abstract}
Protected edge modes are the cornerstone of topological states of matter.
The simplest example is provided by the integer quantum Hall state at Landau level filling unity, 
which should feature a single chiral mode carrying electronic excitations. In the presence 
of a smooth confining potential it was hitherto believed that this picture may only be partially 
modified by the appearance of additional counterpropagating integer-charge modes. Here, we 
demonstrate the breakdown of this paradigm: The system favors the formation of 
edge modes supporting fractional excitations. This accounts for previous observations, and 
leads to additional predictions amenable to experimental tests.
\end{abstract}

\maketitle

\textit{Introduction.} 
Edge modes are responsible for many of the exciting properties of quantum Hall (QH) states~\cite{Halperin1982}:
While the bulk of a QH state is gapped, the edge supports one-dimensional gapless chiral modes~\cite{Wen1990}. 
Although several transport properties of these modes are universal and determined by the topological invariants 
characterizing the bulk state, their detailed structure depends on the interplay between the edge
confining potential, electron-electron interaction, and disorder-induced backscattering. As the confining potential is 
made less steep, the chiral edges of integer~\cite{CSG1992,Dempsey1993,ChamonWen,
Sondhi_PRL_96,KunYangIQHS,Switching2017} and fractional~\cite{MacDonald_PRL_90,Johnson_PRL_91,
MacDonald_JP_93,Meir93,KFP1994,KaneFisher_PRB_95,KunYang_2002,KunYang_2003,KunYang_2008,KunYang_2009,Ganpathy_PRB_03,WMG_PRL_2013} 
QH phases and the helical edges of 
time-reversal-invariant topological insulators~\cite{Yuval2017} may undergo a quantum phase transition 
(or ``edge reconstruction''), while the bulk state remains untouched. Edge reconstruction may be driven by 
charging or exchange effects and leads to a change in the position, ordering, number, and/or nature of the 
edge modes.

Arguably the simplest example is provided by the edge of the $\nu = 1$ QH state. When confined 
by a sharp potential, this state supports a single gapless chiral integer mode with charge $e^{*}=1$; 
the electronic density steeply falls from its bulk value to zero at the edge. Smoothening the confining potential 
and accounting for the incompressibility of QH states leads to the formation of an outer, finite density reconstructed strip. 
Employing a self-consistent Hartree-Fock (HF) scheme, Chamon and Wen~\cite{ChamonWen} found that this additional strip 
can be described as a $\nu = 1$ QH state [Fig.~1(a)]. Such a state allows the local density to assume an integer value, 
leading to a smooth variation of the coarse-grained density from its bulk value to zero. Reconstruction introduces an 
additional pair of counterpropagating gapless chiral modes at the edge. The HF approximation is limited to Slater-determinant 
states, entailing these to be integer modes ($e^{*} = 1$). Exact diagonalization of the $\nu = 1$ phase~\cite{ChamonWen}
(and of fractional phases~\cite{KunYang_2002,KunYang_2003,KunYang_2008,KunYang_2009}) is consistent with the expected 
picture, but is limited to very small systems, rendering it hard to confirm the precise filling factor of the 
side strip or the nature of edge modes.

Recent transport experiments on the $\nu = 1$ state~\cite{Yacoby2012,Heiblum2019} have led to some surprising 
observations regarding the edge structure. Exciting the $\nu = 1$ edge at a quantum point contact (QPC), Ref.~\cite{Yacoby2012} 
observed a flow of energy but not charge upstream from the QPC, possibly indicating the presence of upstream neutral modes. 
Reference~\cite{Heiblum2019} has studied the interference of the edge modes in an electronic Mach-Zehnder interferometer. As the 
bulk filling factor is reduced from 2 to less than 1, reduction in the visibility of the interference pattern has been observed, 
with full suppression for $\nu \leq 1$. This is another indication of the presence of upstream neutral modes~\cite{Moshe_PRL_2016}. 
However, it is inconsistent with Chamon and Wen's picture of only integer-charge modes, which can lead to upstream charge 
propagation, but not to upstream neutral modes. Reference~\cite{Heiblum2019} also found a fractional conductance plateau 
with $g = 1/3 \times e^2/h$ by partially pinching off a QPC in the $\nu = 1$ bulk state. This too is incompatible with the 
edge structure of Fig.~1(a). To cap it all, the conductance plateau observed was accompanied by shot noise with a quantized 
Fano factor 1, which seems to suggest the edge modes do possess an integer charge. 
Fractional modes were also observed at the $\nu = 1$ edge through direct imaging of the local 
density~\cite{Beltram2012,Thomas2014} as well as in recent transport experiments~\cite{Karmakar2020}.

Here, we propose another picture of the reconstructed edge of the $\nu=1$ phase, and show that it accounts for all 
these seemingly contradictory observations. We establish that reconstruction may introduce a different type of counterpropagating modes, 
namely \textit{fractionally charged} ($e^{*} = 1/3$) modes. This is the case when the strip of electrons separated at the edge 
forms a $\nu = 1/3$ Laughlin state [Fig.~1(b)] instead of the commonly assumed $\nu = 1$ state 
(such an edge structure was first suggested in Ref.~\cite{Heiblum2019}). 
To go beyond the constraints 
of the HF approximation [which imply an integer (0 or 1) occupation of each single-particle state], we 
follow the approach by Meir~\cite{Meir93} and treat the two edge configurations depicted in Fig.~1 
as variational states, and compare their respective energies for different strip size ($N_{S}$) and separation ($L_{S}$) 
as a function of the slope of the confining potential. We find that for smooth slopes the fractionally reconstructed edge 
[Fig.~1(b)] is energetically favorable. Our analysis then demonstrates that \textit{fractional} edge reconstruction may be 
much more robust than integer reconstruction.

\begin{figure}[t]
  \centering
  \includegraphics[width=\columnwidth]{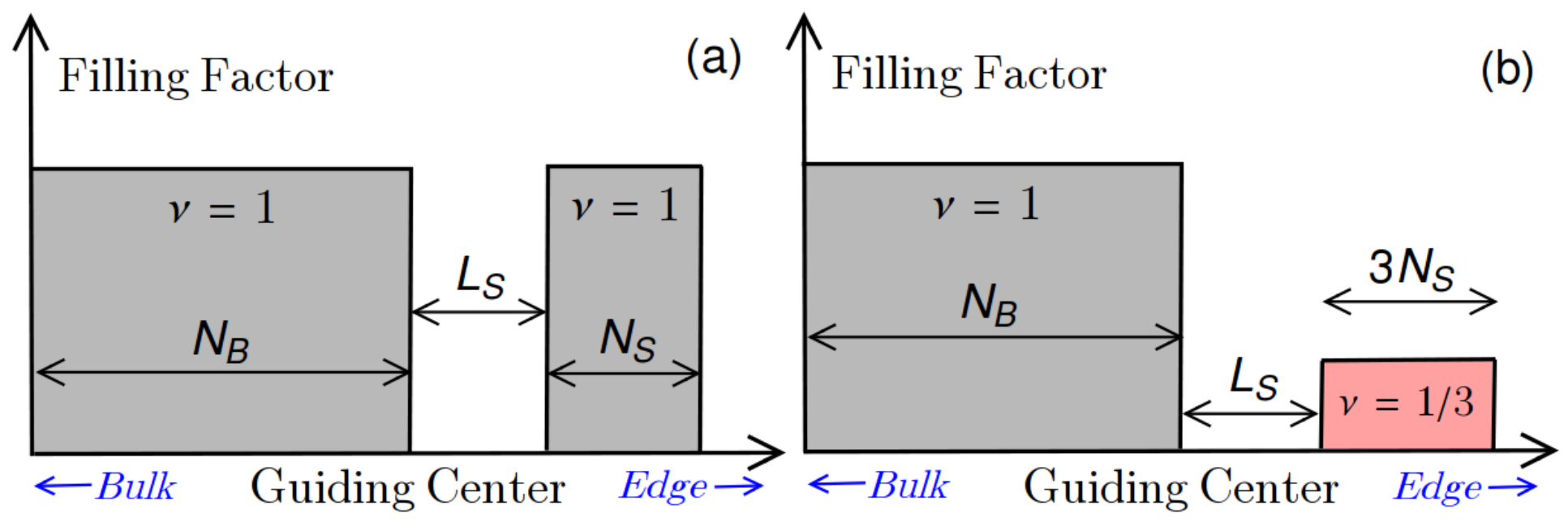}
  \caption{ Schematic representation of two possible configurations at the reconstructed edge of the $\nu=1$ state. 
  Letting the confining potential become smoother, $N_{S}$ electrons may separate from the bulk by $L_{S}$ guiding centers, 
 forming a strip of (a) a $\nu = 1$ state~\cite{ChamonWen} or (b) a $\nu = \frac{1}{3}$ Laughlin state. }
\end{figure}

The intricate edge structure involving a downstream $e^{*} = 1$ mode along with a pair of counterpropagating $e^{*} = 1/3$ 
modes has several experimental consequences. First, with such an edge structure the two-terminal (electrical) conductance 
would vary from $g_{\text{2T}} = e^2/h$ in a long sample (with full edge equilibriation) to $g_{\text{2T}} = 5/3 \times e^2/h$ 
in a short sample (with no equilibration)~\cite{Yuval_AP_2017,Nosiglia2018}. This would be a \textit{smoking gun} signature 
of the edge structure proposed here. Second, in the presence of disorder-induced tunneling and intermode interactions, the 
counterpropagating modes $e^{*} = 1$ and $1/3$ are renormalized to two effective modes of charge $e^{*}_{\ua}$ and 
$e^{*}_{\da}$~\cite{KFP1994,KaneFisher_PRB_95,Yuval_AP_2017} (here, $\ua/\da$ denote the upstream/downstream modes).
When biased, the upstream mode can carry a heat flow, which, in the particularly interesting case of $e^{*}_{\ua} = 0$ 
and $e^{*}_{\da} = 2/3$, may appear without an accompanying upstream charge flow. Such neutral modes have been observed 
in hole-conjugate QH states~\cite{Yacoby2012,Bid2009,Bid2010,Gurman2012,Yaron2012,Inoue2014}. Bias of the neutral modes 
can cause stochastic noise in the charge modes through the generation of quasihole-quasiparticle 
pairs~\cite{Bid2010,Cohen2019,Park2019,Spanslatt2020}. Below we show that this could account for the aforementioned 
Fano factor 1~\cite{Heiblum2019}. Moreover, neutral modes may also lead to suppression of interference in Mach-Zehnder 
interferometers~\cite{Moshe_PRL_2016}, in line with existing experiments.

\textit{Basic setup.} 
We consider a $\nu=1$ state on a disk. In the symmetric gauge, $e\vec{A}/\hbar = (-y/2\ell^2, x/2\ell^2)$,
the wave function of single-particle states in the lowest Landau level are
$ \phm (\vcr\,) = \left( r / \ell \right)^{m} e^{-im\theta_{\text{r}}} 
e^{-\left(\frac{r}{2\ell}\right)^2} / \sqrt{2^{m+1} \pi m! \ell^2} $,
where $(r,\theta_{\text{r}})$ are the polar components of $\vcr$ in the $x$-$y$ plane; 
$\phm$ is an angular momentum eigenfunction with eigenvalue $\hbar m$, centered at $r = \sqrt{2 m} \ell$ where
$\ell$ is the magnetic length. 
Assuming spin-polarized electrons and neglecting higher Landau levels, the Hamiltonian is $H = H_{ee} + H_{c}$,
where $H_{ee}$ is the interaction part while $H_{c}$ is a circularly symmetric one-body confining potential.
Denoting $E_{c} = e^2/\epsilon_0 \ell$, $H_{ee} = (E_{c}/2) \sum_{m_{1},m_{2},n} V_{m_1 m_2 ; n}^{ee} 
c_{m_1 + n}^{\dagger} c_{m_2}^{\dagger} c_{m_2 + n} c_{m_1}$ and
$H_{c} = E_c \sum_{m} V_{m}^{c} c_{m}^{\dagger} c_{m} $, 
where $V^{ee}$ is the two-body Coulomb matrix element and $V^{c}$ is the matrix element of the
confining potential. The total angular momentum $L$ is a good quantum number. The edge confining potential 
reads~\cite{Meir93} 
\begin{align}
V_{c} (r)= \Bigg\{\begin{array}{cc}
0 & r < r_0 - \frac{w \ell}{2},\\
\frac{s}{w \ell}\big(r - r_0 + \frac{w \ell}{2} \big) & r_0 - \frac{w \ell}{2} < r < r_0 + \frac{w \ell}{2},\\
s & r > r_0 + \frac{w \ell}{2}, \end{array}
\end{align}
where $r_0$ is the radius of a compact $\nu = 1$ state. The dimensionless parameter $s$ sets the overall height of the potential, 
which we henceforth fix to $s=7$. The steepness of the potential is controlled by the dimensionless width $w$.

We consider two classes of variational states (shown in Fig.~1), corresponding to an integer 
[Chamon-Wen~\cite{ChamonWen}, Fig.~1(a)] and a fractional [Fig.~1(b)] reconstructed edge. Both are controlled 
by two parameters: the total occupancy $N_S$ of the reconstructed edge strip, and the number $L_S$ of empty 
orbitals separating it from the bulk. The latter contains $N_B$ electrons, such that the total number of electrons 
$N_{S} + N_{B}$ is fixed (to be 100). The Chamon-Wen family of states includes the compact edge configuration 
($N_{S} = 0 = L_{S}$) which is the ground state for sharp confining potentials. For smoother confining potentials, 
the lowest energy state is expected to be at nonzero $N_{S}$ and $L_{S}$. In this case, a comparison of the energies 
of the states in the two classes determines whether fractionally charged modes could appear at the edge of the $\nu=1$ phase.

\textit{Variational ansatz: Integer edges.---} Figure~1(a) represents a Slater-determinant state of $N_{S} + N_{B}$ 
electrons. It can be written as $|N_{B},0\rangle \otimes |N_{S},N_{B} + L_{S}\rangle$,
where 
\begin{align}
  |N,L\rangle = c_{L+N-1}^{\dagger} \, c_{L+N-2}^{\dagger} \, \ldots \, c_{L+1}^{\dagger} \, c_{L}^{\dagger} |0\rangle.
\end{align}
The energy and angular momentum of each state in the integer class of reconstructions can be found easily once 
the Coulomb matrix elements are 
known~\cite{Supplemental}.%,Tsiper2002,JainCF,MacDonald93,Laughlin83,Metropolis53}.

\textit{Variational ansatz: Fractional edges.} Figure~1(b) represents the product state of a Slater determinant 
($|N_{B},0\rangle$) with an annulus of the $\nu=1/3$ Laughlin state, containing $N_S$ electrons starting at the guiding center 
$m=N_B+L_S$. The (unnormalized) wave function corresponding to the annulus is
\begin{align} \label{eq:wvfn}
  \prod_{i=1}^{N_{S}} \bigg[ z_i ^{N_{B} + L_{S}} \bigg]
  \bigg[ \prod_{i < j} \big( z_i - z_j \big)^{3} \bigg] e^{-\frac{1}{4} \sum_{i} |z_{i}|^2},
\end{align}
where $z_n = x_n - i y_n$ is the coordinate of the $n$th particle. The energy and angular momentum of states in 
this class involve the Coulomb energy and average occupations of the Laughlin state [Eq.~(\ref{eq:wvfn})]. We evaluate 
these using standard classical Monte Carlo techniques~\cite{Supplemental}. 

\begin{figure}[t]
  \centering
  \includegraphics[width=\columnwidth]{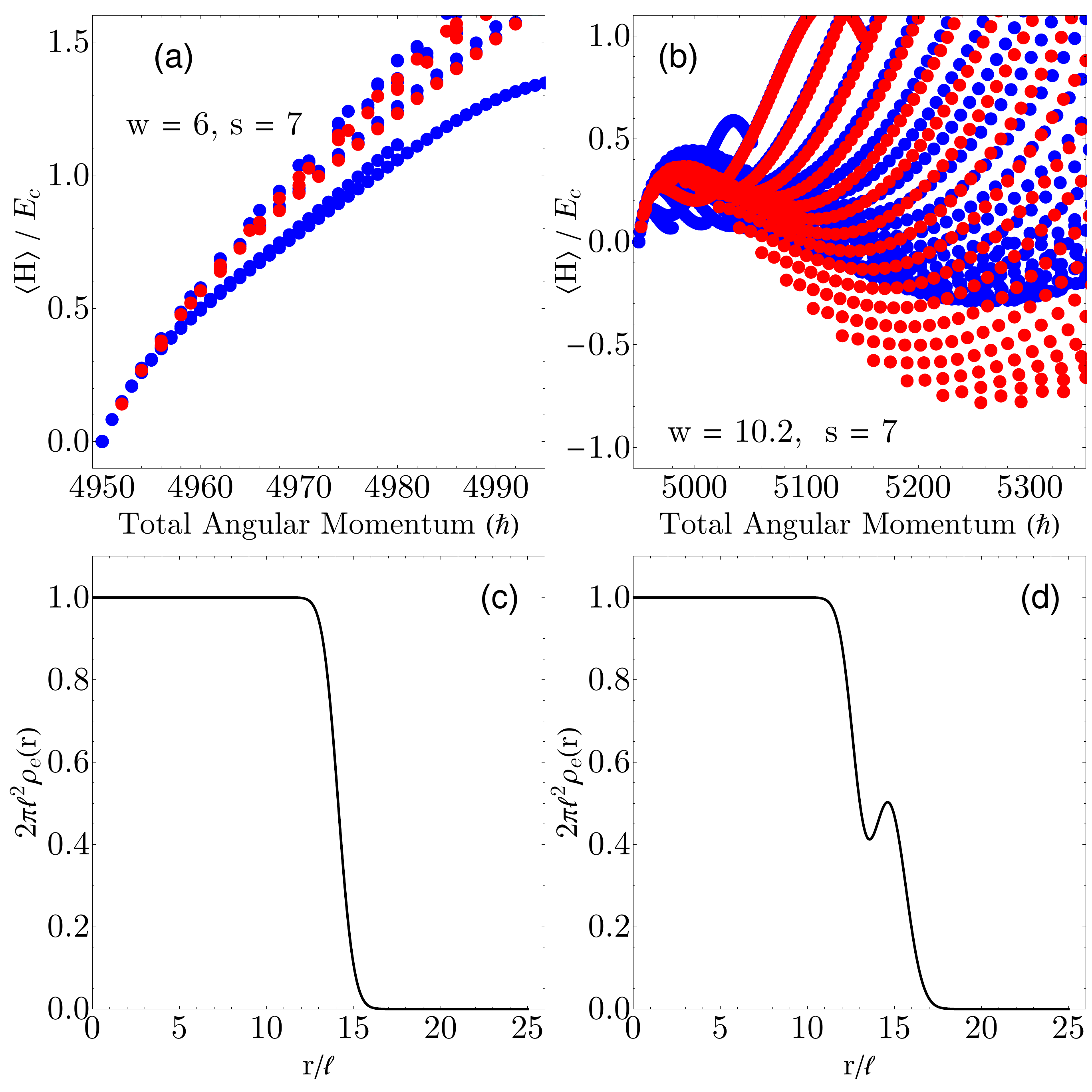}
  \caption{ Variational analysis for $N_{S} + N_{B} = 100$ and $s = 7$. (a), (b) The energy of the two variational 
  states as a function of the total angular momentum at (a) $w = 6.0$ and (b) $ w = 10.2$. The energy of the unreconstructed 
  state has been subtracted to make comparison easier. The blue (red) dots correspond to states with $\nu = 1$ 
  ($\nu = \frac{1}{3}$) reconstruction at the edge. For sharp edges ($w < 10$) the ground state is the one with minimum 
  angular momentum, implying that $L_{S} = 0$, hence no edge reconstruction. In this case, we expect a single downstream edge 
  mode supporting $e^{*} = 1$ quasiparticles. For smooth edges ($w > 10$) the ground state shifts to a higher angular momentum
  sector implying that the electronic disk expands and the edge undergoes reconstruction. (b) shows that 
  a fractional reconstruction is energetically favorable to an integer reconstruction. This is true for all $w > 10$. 
  Thus the reconstructed edge supports counterpropagating modes with fractional charges. (c) and (d) depict the electronic densities 
  of the ground state at (c) $w = 6.0$ and (d) $w = 10.2$. The nonmonotonic variation of density at the edge is another 
  signature of the presence of additional emergent modes.
}
\end{figure}

\textit{Results.} Figure~2 shows the total energies and the ground state densities for the two class of variational states at different 
confining potentials. In Figs.~2(a) and~2(b) the blue dots correspond to integer edges while the red dots 
correspond to the fractional edge states. For a sharp confining potential [$w < 10$, Fig.~2(a)] the lowest energy 
state is the one with the minimal angular momentum (in this case $4950\hbar$). This corresponds to the 
unreconstructed $\nu = 1$ state with a single chiral edge mode. Figure~2(c) shows the electronic 
density in this case, which drops monotonically from $1/2\pi\ell^2$ to 0. 

For smoother potentials [$w > 10$, Fig.~2(b)] the lowest energy state has a much larger angular momentum
($ 5256 \hbar$ for $w=10.2$ with $N_{S} = 18$ and $L_{S} = 0$) than the compact state. 
Correspondingly, Fig.~2(d) shows that the density varies nonmonotonically at the edge~\cite{fnoteLs}. 
The states with a fractional edge are found to have a lower energy than the states with an integer edge 
\textit{whenever reconstruction is favored}~\cite{Note2by3}. This is the main result of this work. 
We have verified that it does not depend on the precise form of the confining potential~\cite{Supplemental}. 
We now turn to 
discuss the experimental consequences of such a reconstruction and compare them to the observations reported 
in literature so far.

\begin{figure}[t]
  \centering
  \includegraphics[width=\columnwidth]{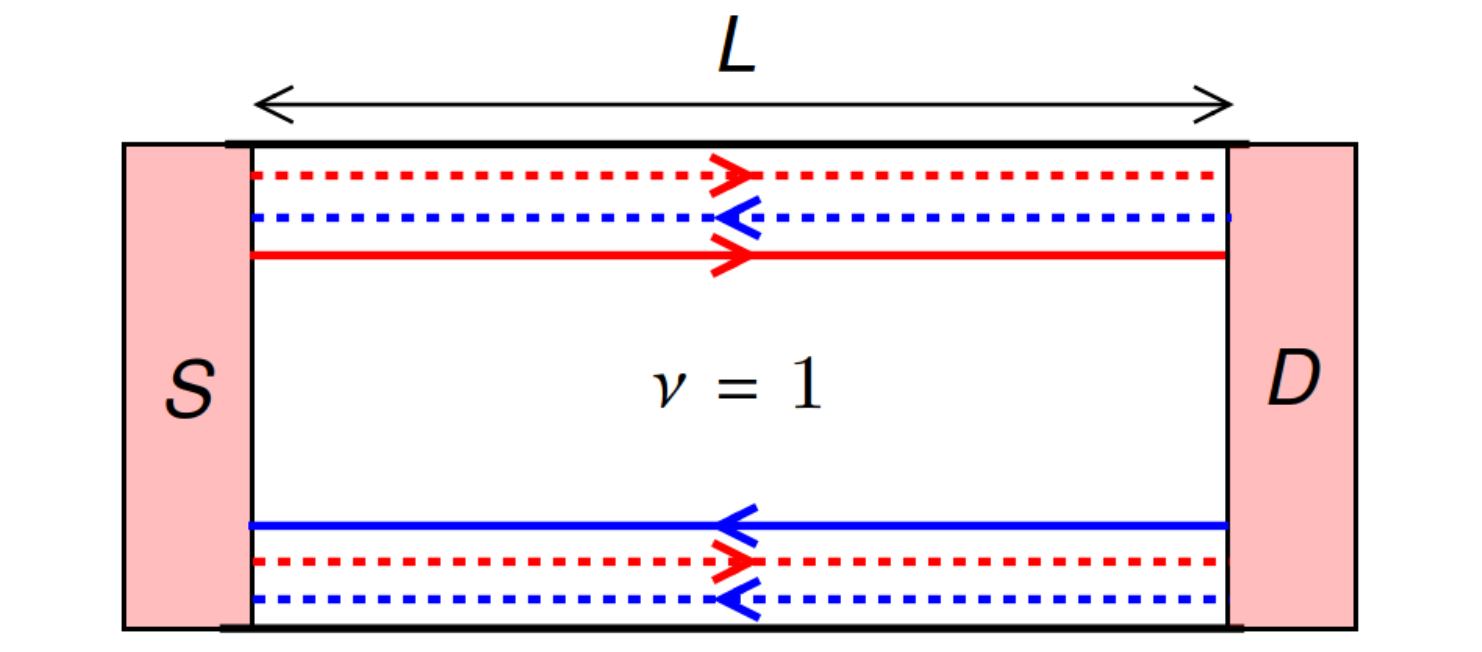}
  \caption{ Two-terminal transport experiment at $\nu = 1$, with an edge structure as calculated for a 
  disk geometry (cf. text). The solid (dashed) lines indicate the integer (fractional) chirals at the two 
  edges of the sample. The red (blue) chirals are biased (unbiased) due to the source S (drain D). For 
  $L \ll \ell_{\text{eq}}$ ($\ell_{\text{eq}}$ is the intermode equilibriation length) the conductance is 
  $g = 5/3 \times e^2/h$ ($3 \times e^2/h$) for fractional (integer) edge reconstruction [cf. Figs. 1(b) and 1(a)].
  For a fully equilibrated edge ($L \gg \ell_{\text{eq}}$), the conductance reduces to $g = e^2/h$ in 
  both cases, as expected for the unreconstructed $\nu = 1$ state. } 
\end{figure}

\textit{Two-terminal conductance.} Let us consider the setup shown in Fig.~3, where the edge structure is based on our analysis of a disk geometry. 
The chiral modes emanating from the source (S) are biased with respect to those emerging from the drain (D). 
Due to disorder-induced intermode tunneling, the counterpropagating chirals at each edge will equilibrate over 
a typical length $\ell_{\text{eq}}$. For a fully equilibrated edge ($ L \gg \ell_{\text{eq}}$), the two-terminal 
conductance is $e^2/h$, as expected for the $\nu = 1$ QH state. Note that this would be the case for both
sharp and smooth edges and for both integer and fractional reconstructions. 

For $ L \ll \ell_{\text{eq}}$, the detailed structure of the edge underlies the conductance.  
For a sharp edge transport takes place through a single integer chiral, hence the electric conductance would 
retain the values $e^2/h$. This is different for smooth edges. The electric conductance is sensitive to the 
number as well as the nature of the modes; with a pair of counterpropagating fractional edges, the electric 
conductance becomes $5/3 \times e^2/h$~\cite{Yuval_AP_2017,Nosiglia2018}. Such an observation would uniquely 
identify the edge structure proposed here [Fig.~1(b)]---a smoking gun signature of fractional edge 
reconstruction~\cite{ThermalTransport}.

\textit{Neutral modes.} Consider the fractional reconstruction of Fig.~1(b). Labeling the outermost 
channel as $1$ and the innermost edge as $3$ [cf. Fig.~4(a)], the low energy dynamics of the three modes is 
described by three chiral bosonic fields $\phi_j$ ($j = 1,2,3$) satisfying the Kac-Moody algebra, 
$[\phi_{j_1} (x), \phi_{j_2} (x')] = i \pi \big[ K^{-1} \big]_{j_1 , j_2} \text{sgn}(x - x')$, 
where the $K$ matrix is diagonal with $K_{1,1} = 3, K_{2,2} = -3, K_{3,3} = 1$.
The inner two modes are counterpropagating charge modes of $\nu = 1$ and $\nu = 1/3$ type. This is
precisely the edge structure of the hole-conjugate $\nu = 2/3$ FQH state. Since $L_{S}$ is typically small~\cite{fnoteLs}, 
in the presence of disorder-induced backscattering and interactions the two charge modes can hybridize [Fig.~4(a)], 
resulting in a downstream charged mode $\phi_{c}$ and an upstream neutral mode $\phi_{n} $, 
which are effectively decoupled at low energies~\cite{KFP1994}. This $K$ matrix is diagonal with 
$K_{1,1} = 3, K_{c, c} = 1, K_{n, n} = -1$. 
We note that here the outermost mode ($\phi_1$) is kept untouched (cf. Fig.~4). 

\begin{figure}[t]
  \centering
  \includegraphics[width=\columnwidth]{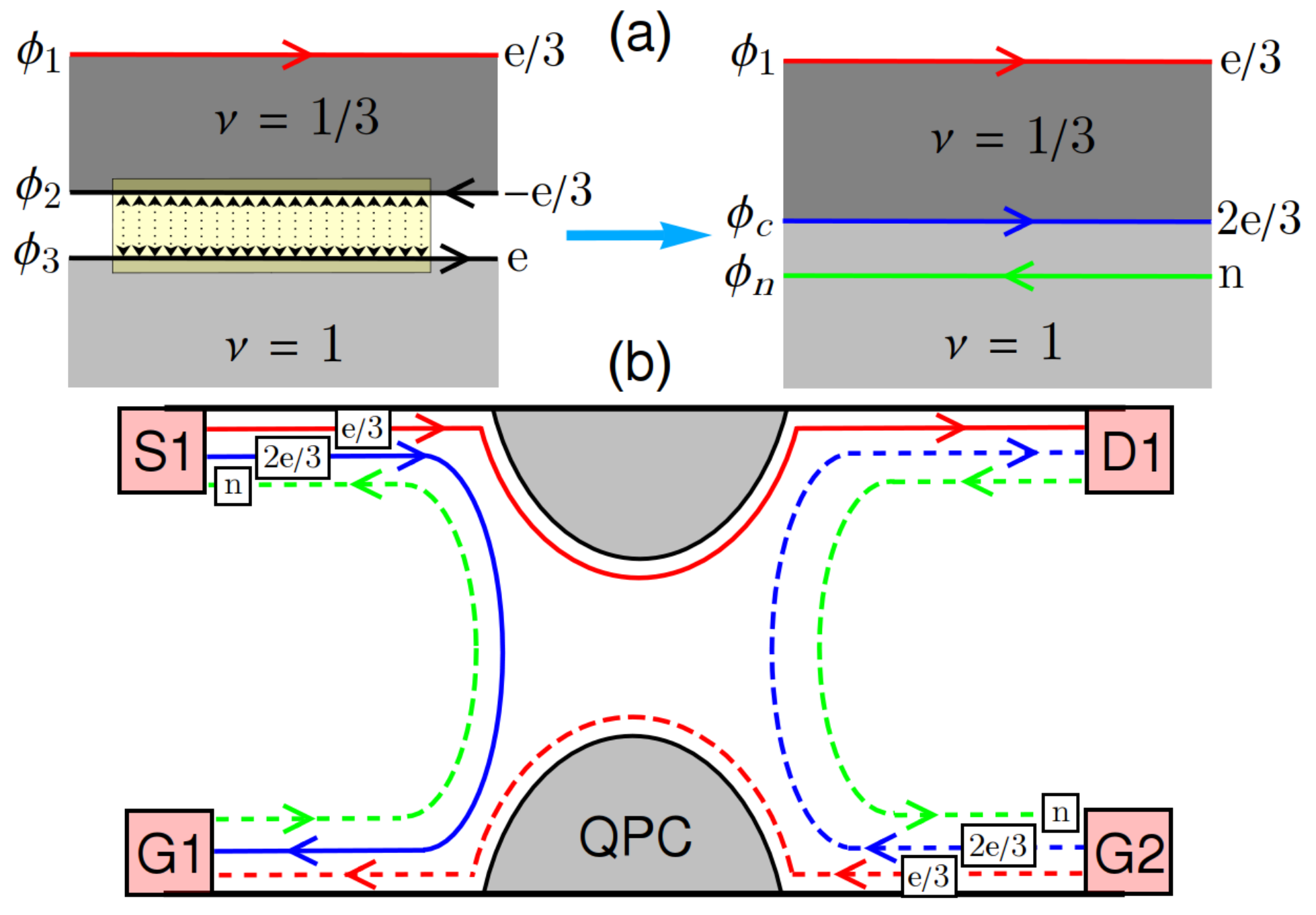}
  \caption{ (a) Renormalization of the inner two edge modes due to interactions and disorder-induced
  backscattering into a downstream charge ($\phi_{c}$) and upstream neutral ($\phi_{n}$)
  mode. Only the inner two modes are assumed to couple, since within the variational 
  calculation, the width of the $\nu = 1/3$ strip increases as the edge potential is made smoother 
  but the separation between the $\nu = 1$ and $\nu = 1/3$ regions remains constant. Thus the 
  outermost edge mode ($\phi_1$) can be assumed to be physically separated from the inner two modes 
  ($\phi_{2,3}$)~\cite{WMG_PRL_2013}. (b) A single QPC tuned to the transmission plateau $t = 1/3$. 
  The bulk on both sides of the QPC is in the $\nu = 1$ state with a reconstructed and renormalized 
  edge. Solid (dashed) lines correspond to biased (unbiased) modes. }
\end{figure}

The experimental consequences of this emergent neutral mode are similar to the neutral modes in 
hole-conjugate states. For instance, it can lead to an upstream thermal current, which was reported 
in Ref.~\cite{Yacoby2012}, accompanied by an upstream shot noise (see below)~\cite{Sabo2017,Spanslatt2019}. The presence
of the neutral mode can also hinder observation of interference effects in Mach-Zehnder setups~\cite{Moshe_PRL_2016} 
as reported in Ref.~\cite{Heiblum2019}. 

\textit{Fractional conductance plateau and noise.} The presence of fractionally 
charged chiral modes at the edge has clear experimental consequences for 
transport measurements. Consider for example the single QPC setup of Fig.~4(b). Here, the bulk filling 
factor is $\nu = 1$ and the current is transmitted from the
source (S1) to the drain (D1). When the QPC is fully open then the conductance would be $e^2/h$, 
as expected from the bulk topological index. However, due to the edge structure discussed above, 
it is also possible to pinch off the QPC, so that only the outermost mode ($\phi_1$) is transmitted 
while the inner two modes are completely reflected. In this case there would be a fractional 
conductance plateau at $1/3 \times e^2/h$ while the bulk filling factor remains 1. Such a plateau 
was reported in Ref.~\cite{Heiblum2019}.

Interestingly, although the conductance is quantized, the system could exhibit shot noise
on the conductance plateau. Under the assumption of coherent propagation of the neutral mode, and
provided certain symmetry conditions are satisfied~\cite{Cohen2019,Jinhong2020}, the Fano factor is quantized.
Such a quantized noise at the $1/3$ conductance plateau has been reported in Ref.~\cite{Heiblum2019}.
Below we sketch the underlying physics relying on our fractionally reconstructed edge picture. 

Consider the setup shown in Fig.~4(b). The source S1 on the upper left-hand side of the QPC biases both 
charge modes emanating from it with the same voltage (say $V$). The current in the two modes is 
$I_{1} = V/3 \times e^2/h$, $I_{c} = 2V/3 \times e^2/h$, and the total current is thus $I = I_1 + 
I_{c} = V \times e^2/h$. The current ($I_i$, $i = 1,c$) in a given mode is related to the corresponding quasiparticle 
density ($n_i$) through $I_{1} = e/3 \times v_{1} n_{1}$ and $I_{c} = 2e/3 \times v_{c} n_{c}$, 
where $v_{i}$ are the corresponding velocities, implying $v_{1} n_{1} = v_{c} n_{c}$. Therefore if $N$ 
quasiparticles of charge $\frac{1}{3}$ emanate from the S1 in time $\tau$, then $N$ quasiparticles of charge 
$\frac{2}{3}$ also emanate in the same time interval. The total current ($I$) is $I = e/3 \times N/\tau + 
2e/3 \times N/\tau = eN/\tau$.

Now, on the upper right-hand side of the QPC, the outermost $e/3$ mode is biased while the inner 
$2e/3$ mode is grounded, and therefore the two modes will equilibrate through tunneling processes,
which would also create excitations in the neutral mode. If there were $N$ quasiparticles
in $\phi_{1}$, then after equilibration with $\phi_{c}$ there would be $N/3$ quasiparticles left in 
both charged modes and $2N/3$ neutral excitations in the upstream neutral mode. 
These neutral excitations would move to the lower right-hand side of the QPC and decay into 
quasiparticle-quasihole pairs in the charge modes. This generates stochastic noise in the charged modes 
because each decay process can randomly generate either a quasiparticle (quasihole) in the outermost 
(inner) mode or vice versa. This decay process would lead to a stochastic tunneling 
of $N/3$ electronic excitations into $\phi_{c}$, which eventually reach the drain D1. Similarly, 
on the lower left-hand side of the QPC, a biased $2e/3$ mode flows in parallel to an unbiased $e/3$ mode. 
Their mutual equilibration would again generate $2N/3$ neutral excitations. These decay on the upper 
left-hand side of the QPC and generate $2N/3$ excitations in the $\phi_{1}$ mode entering the drain D1.

As a result of the above, the charge entering the drain in time $\tau$ is $ Q = e/3 \times N/3 + 2e/3 \times N/3 + 
e/3 \times \sum_{i = 1}^{2N/3} a_i + 2e/3 \times \sum_{i = 1}^{N/3} b_i$, where $a_i$ and $b_i$ are 
random variables which take values $\pm 1$ with equal probability, and describe the noise generated in the modes due to 
the neutral excitation decay described above. This implies that the average current arriving at the drain is 
$I_{D} = \langle Q \rangle/\tau = eN/3 = I/3$ (consistent with a transmission of $1/3$). The variance of the 
charge is $\delta Q^2 = \langle Q^2 \rangle - \langle Q \rangle^2 = e^2/9 \times \sum_{i = 1}^{2N/3} a_i^2 + 
4e^2/9 \times \sum_{i = 1}^{N/3} b_i^2 = 2Ne^2/9 = 2e/9 \times I\tau$. The effective
Fano factor is $F_{\text{eff}} = \delta Q^2 /I \tau \times 1/e t(1-t)$. 
Using $t = 1/3$ we obtain $F_{\text{eff}} = 1$, which coincides with the observation of Ref.~\cite{Heiblum2019}.

{\it Conclusions.} We have studied edge reconstruction at the boundary of $\nu = 1$ integer 
quantum Hall state. Previously reported Hartree-Fock calculations show that upon smoothening the confining 
potential a new strip of $\nu = 1$ QH state is formed at the edge, introducing counterpropagating \textit{integer} 
modes~\cite{ChamonWen}. Going beyond the mean-field approximation, we have performed a variational calculation, 
where we have compared the above ansatz to a new one, in which the electronic strip forms a $\nu = 1/3$ Laughlin 
state. We have found that such fractional reconstruction is always energetically favorable, implying that 
\textit{fractional modes can appear at the boundary of integer QH states}. We have discussed the experimental 
consequences of such a fractionally reconstructed edge, which nicely square with previous  
measurements, and provide predictions for future experiments. Our finding sets the stage for a future detailed 
investigation of coherent as well as incoherent transport in designed geometries, implementing the idea of 
fractionally reconstructed edges.

\begin{acknowledgments}
We acknowledge useful discussions with M. Heiblum and J. Park. U.K. was supported by the Raymond and Beverly
Sackler Faculty of Exact Sciences at Tel Aviv University and by the Raymond and Beverly Sackler Center for
Computational Molecular and Material Science. M.G. and Y.G. were supported by the Israel Ministry of Science and 
Technology (Contract No.~3-12419).~M.G. was also supported by the Israel Science Foundation (ISF, Grant No.~227/15) 
and US-Israel Binational Science Foundation (BSF, Grant No.~2016224).~Y.G. was also supported by CRC~183 
(project~C01), the Minerva Foundation, DFG Grant No.~RO~2247/8-1, DFG Grant No.~MI~658/10-1, and the 
GIF Grant No.~I-1505-303.10/2019. 
\end{acknowledgments}

\onecolumngrid
\clearpage

\setcounter{affil}{0}
\setcounter{page}{1}
\renewcommand{\thefigure}{S\arabic{figure}}
\setcounter{figure}{0}
\renewcommand{\theequation}{S\arabic{equation}}
\setcounter{equation}{0}
\renewcommand\thesection{S\arabic{section}}
\setcounter{section}{0}

\title{Supplemental material for ``Fractional Edge Reconstruction in Integer Quantum Hall Phases''}
\author{Udit Khanna}
\affiliation{Raymond and Beverly Sackler School of Physics and Astronomy, Tel-Aviv University, Tel Aviv, 6997801, Israel}
\affiliation{Department of Condensed Matter Physics, Weizmann Institute of Science, Rehovot 76100, Israel}
\author{Moshe Goldstein}
\affiliation{Raymond and Beverly Sackler School of Physics and Astronomy, Tel-Aviv University, Tel Aviv, 6997801, Israel}
\author{Yuval Gefen}
\affiliation{Department of Condensed Matter Physics, Weizmann Institute of Science, Rehovot 76100, Israel}

\date{\today}

\begin{abstract}
This supplemental material provides details regarding our numerical analysis as well as 
extensions of our analysis. Sections I and II describe the variational method used to find the lowest energy 
state for integer (Section I) and fractional (Section II) edge reconstruction. Section III presents results 
of our variational analysis, employing a different confining potential. Section IV summarizes an exact 
diagonalization analysis of the same setup.
\end{abstract}

\maketitle

\section{I.\,\,\,\,\,\,      Integer Reconstruction}

Fig.~1(a) represents a Slater determinant of $N_{S} + N_{B}$ 
electrons. For convenience, we write it as the product of two Slater determinants, $|N_{B},0\rangle \otimes 
|N_{S},N_{B} + L_{S}\rangle$ where
\begin{align}
  |N,L\rangle = c_{L+N-1}^{\dagger} \, c_{L+N-2}^{\dagger} \, \ldots \, c_{L+1}^{\dagger} \, c_{L}^{\dagger} |0\rangle.
\end{align}

The total angular momentum (in units of $\hbar$) of $|N,L\rangle$ is $NL + N (N-1)/2$, and that 
of the combined state is just the sum of the angular momenta of its two components 
\begin{align}
  N_{S} L_{S} + \frac{1}{2} (N_{B} + N_{S}) (N_{B} + N_{S} - 1). 
\end{align}
The second term above is the angular momentum of the compact state ($L_{S} = 0$). Thus the 
unreconstructed state has the smallest possible angular momentum for a fixed number of electrons 
($N_{S} + N_{B}$) in the lowest Landau level. We have used $N_{S} + N_{B} = 100$, which corresponds 
to minimum angular momentum 4950 ($\hbar$). 

The energy of $|N,L\rangle$ is $\langle N,L | H_{ee} | N,L \rangle + \langle N,L | H_{c} | N,L \rangle$ 
where,
\begin{align} \label{EneSla}
  \langle N,L | H_{ee} | N,L \rangle &= E_{c} \sum_{\substack{i,j = L \\ (i < j)}}^{N+L-1} 
    \bigg( V_{i j ; 0}^{ee} - V_{i i ; j-i}^{ee} \bigg), \\ \label{EneSlc}
  \langle N,L | H_{c} | N,L \rangle &= E_c \sum_{i = L}^{N+L-1} V_{i}^{c}.
\end{align}
The energy of the full state consists of the sum of the energies of its constituents, as well as their 
two-body interaction energy,
\begin{align}
  E_{c} \sum_{i = 0}^{N_{B}-1} \sum_{j = N_{B} + L_{S}}^{N_{B} + L_{S} + N_{S} - 1} 
    \bigg( V_{i j ; 0}^{ee} - V_{i i ; j-i}^{ee} \bigg).
\end{align}
Therefore, the energy and angular momentum of each state in the integer class of reconstructions can be 
computed easily once the matrix elements are known. In the disk geometry, the Coulomb matrix elements for 
lowest Landau level states can be found analytically~\cite{STsiper2002,SJainCF}. The matrix elements of confining 
potentials are given by, 
\begin{align}
  V_{m}^{c} &= \int d^{2}r \, V_{c} (r) |\phi_{m} (\vcr\,)|^2
\end{align}

We note that for sharp and moderately smooth confining potentials 
($w \leq 14$) and in the absence of Landau level and spin mixing, the minimum energy state within this class of 
reconstructions is precisely the ground state in the self-consistent Hartree-Fock (HF) approximation.

\section{II.\,\,\,\,\,\,      Fractional Reconstruction}

Fig.~1(b) represents the product state of a Slater determinant ($|N_{B},0\rangle$) 
with an annulus of the $\nu=1/3$ Laughlin state ($| \Psi_{\frac{1}{3}} \rangle $), 
containing $N_S$ electrons starting at the guiding center $m=N_B+L_S$. The (unnormalized)
wavefunction corresponding to $| \Psi_{\frac{1}{3}} \rangle $ is,
\begin{align}
  \prod_{i=1}^{N_{S}} \bigg[ z_i ^{N_{B} + L_{S}} \bigg]
  \bigg[ \prod_{i < j} \big( z_i - z_j \big)^{3} \bigg] e^{-\frac{1}{4} \sum_{i} |z_{i}|^2},
\end{align}
where $z_i = (x_i - i y_i)/\ell$ is the coordinate of the $i$th particle. 

The angular momentum of the (standard) Laughlin state with $N_{S}$ particles is $\frac{3}{2} N_{S} (N_{S}-1)$. 
Adding $N_{B} + L_{S}$ holes in the center increases the angular momentum by $N_{S} (N_{B} + L_{S})$. 
Then the combined state has a total angular momentum $N_{S} ( L_{S} + N_{S} - 1)
+ \frac{1}{2} (N_{B} + N_{S}) (N_{B} + N_{S}-1)$. Comparing this expression with that 
of the corresponding integer-edge state, we note that this is larger by $N_{S} (N_{S}-1)$. This indicates
that the electronic density of the fractionally reconstructed state varies much more smoothly than the
corresponding integer reconstructed state.

The energy of the combined state is the sum of the energy of the two components (the $\nu = 1$ bulk and
the $\nu = 1/3$ annulus) and their mutual interaction energy. The energy of $| \Psi_{\frac{1}{3}} \rangle$ is  
$ \big[ \langle \Psi_{\frac{1}{3}} | H_{ee} | \Psi_{\frac{1}{3}} \rangle + \langle \Psi_{\frac{1}{3}} 
| H_{c} | \Psi_{\frac{1}{3}} \rangle \big] / \langle \Psi_{\frac{1}{3}} | \Psi_{\frac{1}{3}} \rangle$ where
\begin{align} \label{EneLla}
  \langle \Psi_{\frac{1}{3}} | \Psi_{\frac{1}{3}} \rangle &= \int \prod_{i} d^2 r_i \big| 
  \Psi_{\frac{1}{3}} \big|^2 , \\
  \langle \Psi_{\frac{1}{3}} | H_{ee} | \Psi_{\frac{1}{3}} \rangle &= 
  \int \prod_{i} d^2 r_i \big| \Psi_{\frac{1}{3}} \big|^2 
  \bigg[ \sum_{i < j} \frac{E_{c} \ell}{|\vcr_i - \vcr_j|} \bigg], \\ \label{EneLlc}
  \langle \Psi_{\frac{1}{3}} | H_{c} | \Psi_{\frac{1}{3}} \rangle &= E_{c}
  \sum_{m} \langle \Psi_{\frac{1}{3}} | c_{m}^{\dagger} c_{m} | \Psi_{\frac{1}{3}} \rangle V_{m}^{c} ,
\end{align}
and its interaction energy with the bulk $\nu=1$ state is 
\begin{align}
  E_{c} \sum_{i = 0}^{N_{B}-1} \sum_{j = N_{B} + L_{S}}^{N_{B} + L_{S} + 3N_{S}-3} 
  \frac{\langle \Psi_{\frac{1}{3}} | c_{j}^{\dagger} c_{j}^{} | \Psi_{\frac{1}{3}} 
  \rangle}{\langle \Psi_{\frac{1}{3}} | \Psi_{\frac{1}{3}} \rangle}
  \bigg( V_{i j ; 0}^{ee} - V_{i i ; j-i}^{ee} \bigg).
\end{align}
These expressions involve the Coulomb energy and average occupations of the Laughlin states, 
which we evaluate using standard classical Monte-Carlo techniques~\cite{SJainCF,SMeir93,SMacDonald93} briefly
described below.

\subsubsection*{Coulomb Energy}

The Coulomb energy of $| \Psi_{\frac{1}{3}} \rangle$ is
\begin{align} \label{eq:Coulomb}
  \frac{1}{\int \prod_{i} d^2 r_i \big| \Psi_{\frac{1}{3}} \big|^2} \int \prod_{i} d^2 r_i 
  \big| \Psi_{\frac{1}{3}} \big|^2 \bigg[ \sum_{i < j} \frac{E_{c} \ell}{|\vcr_i - \vcr_j|} \bigg].
\end{align}
Since $|\Psi_{\frac{1}{3}}|^2$ is real and positive, it can be interpreted as a (unnormalized) classical 
probability distribution~\cite{SLaughlin83}. Writing $|\Psi_{\frac{1}{3}}|^2$ as a Boltzmann distribution 
$e^{-\beta U}$, we can make this interpretation concrete by recognizing $U$ as the potential for a 
two-dimensional plasma of charged particles in presence of an impurity of charge $N_{B} + L_{S}$ at the origin. 
The Coulomb energy can then be computed using standard Metropolis sampling~\cite{SMetropolis53}. 

\subsubsection*{Average Occupation}

The average occupation of $m^{\text{th}}$ single-particle state in $| \Psi_{\frac{1}{3}} \rangle$ is
\begin{align} \nonumber
  \langle c_{m}^{\dagger} c_{m} &\rangle_{1/3} = \frac{\langle \Psi_{\frac{1}{3}} | c_{m}^{\dagger} 
  c_{m} | \Psi_{\frac{1}{3}} \rangle}{\langle \Psi_{\frac{1}{3}} | \Psi_{\frac{1}{3}} \rangle} \\ 
  &= \int d^2 r_{1} \, d^2 r_{2} \, \rho_{\frac{1}{3}} (\vec{r}_1, \vec{r}_2) \phi_{m}^{*} (\vec{r}_{1}\,) 
  \phi_{m} (\vec{r}_{2}), 
\end{align}
where $\rho_{\frac{1}{3}}$ is the one-particle density matrix of $| \Psi_{\frac{1}{3}} \rangle$,
\begin{align} \label{eq:DM}
  \rho_{\frac{1}{3}} (\vec{r}_a, &\vec{r}_b) = \frac{N_{S}}{\int \prod_{i} d^2 r_i \big| \Psi_{\frac{1}{3}} \big|^2} 
  \times \\ \nonumber
  &\int \prod_{i=2}^{N_{S}} d^2 r_i \Psi_{\frac{1}{3}} (\vec{r}_{a}, \vcr_{2}, \cdots) \Psi_{\frac{1}{3}}^{*} (\vec{r}_{b}, \vcr_{2}, \cdots).
\end{align}
Computing $\rho_{\frac{1}{3}}$ for all $\vcr_{a}$ and $\vcr_{b}$ using the above expression is very costly. 
To simplify the calculation, we note that both $\phi_m$ and $\Psi_{\frac{1}{3}}$ are eigenstates of the 
angular-momentum operator. Therefore the one-particle density matrix also satisfies
\begin{align}
  \rho_{\frac{1}{3}} (\vec{r}_{a}, &\vec{r}_{b}) = \sum_{m} \langle c_{m}^{\dagger} c_{m} \rangle_{1/3} 
  \phi_m (\vec{r}_{a}) \phi_m^{*} (\vec{r}_{b}) .
\end{align}
In the special case of $\vcr_{b} = r e^{i \theta_{r}}$ and $\vcr_{a} = r e^{i \theta_{r} + i \theta}$, the above 
expression reduces to 
\begin{align}
  \rho_{\frac{1}{3}} (\vcr_{b} , \theta; \vcr_{b}) = \sum_{m} \langle c_{m}^{\dagger} c_{m} \rangle_{1/3} 
  | \phi_m (\vcr_{b}) |^2 e^{-i m \theta} .
\end{align}
Since $\langle c_{m}^{\dagger} c_{m} \rangle_{1/3}$ is non-zero over a contiguous, finite and known range of $m$ 
[namely from $m = N_{B} + L_{S}$ to $m = N_{B} + L_{S} + 3(N_{S} - 1)$], the summation over $m$ can be restricted
to this range without any error. Then we may interpret the above relation as a discrete Fourier transform 
from $m$ to its conjugate $\theta$~\cite{SMacDonald93}. Inverting the Fourier transform we get
\begin{align} \label{eq:Occp}
  \langle c_{m}^{\dagger} c_{m} \rangle_{1/3} | \phi_m (\vcr \,) |^2 &= 
  \frac{1}{3 (N_{S} - 1) + 1} \times \\ \nonumber 
  &\sum_{j = 0}^{3 (N_{S} - 1)} e^{i m \theta_j} \rho_{\frac{1}{3}} (\vcr, \theta_{j}; \vcr\,) ,
\end{align}
where $\theta_{j} = 2 \pi j/[3 (N_{S} - 1) + 1]$. Note that Eq.~(\ref{eq:Occp}) is only true for 
$N_{B} + L_{S} \leq m \leq N_{B} + L_{S} + 3(N_{S} - 1)$. In principle Eq.~(\ref{eq:Occp}) is valid for 
any value of $r$, but in practice the statistical error is minimum when $r \sim \sqrt{2 m} \ell$~\cite{SMacDonald93}. 
Since for large $m$, $|\phi_{m}|^2$ is very sharply peaked at this value of $r$, in this work we evaluate 
the occupation by integrating Eq.~(\ref{eq:Occp}) over $\vec{r}$ to get, 
\begin{align} 
  \langle c_{m}^{\dagger} c_{m} \rangle_{1/3} &= 
  \frac{1}{3 (N_{S} - 1) + 1} \sum_{j = 0}^{3 (N_{S} - 1)} e^{i m \theta_j} \rho_{j}, \\
  \text{where } \rho_{j} &= 
  \int d^2 r \, \rho_{\frac{1}{3}} (\vcr, \theta_{j}; \vcr\,) .
\end{align}
Note that $\theta_j$ is not being integrated over in the previous expression. Then the occupation at any 
$m$ (within the appropriate range) can be found after we evaluate $\rho_j$ for all $j = 0, \cdots, 3 (N_{S} - 1)$. 
Using Eq.~(\ref{eq:DM}) we have,
\begin{align} 
  \rho_j = &\frac{N_{S}}{\int \prod_{i} d^2 r_i  \big| 
  \Psi_{\frac{1}{3}} \big|^2} \times \\ \nonumber
  &\int \prod_{i=1}^{N_{S}} d^2 r_i \Psi_{\frac{1}{3}} (\vec{r}_{1} e^{i \theta_j}, \vcr_{2}, \cdots) 
  \Psi_{\frac{1}{3}}^{*} (\vec{r}_{1}, \vcr_{2}, \cdots).
\end{align}
From the definition of $\Psi_{\frac{1}{3}}$ we obtain
\begin{align}
  &\Psi_{\frac{1}{3}} (\vec{r}_{1} e^{i \theta_j}, \vcr_{2}, \cdots) = \Psi_{\frac{1}{3}} 
  (\{ \vec{r}_{i} \}) \times \mathcal{Z}_{1} (\theta_{j}; \{ \vec{r}_{i} \}), \\
  &\mathcal{Z}_{a} (\theta_{j}; \{ \vec{r}_{i} \}) = 
  e^{-i \theta_{j} (N_{B} + L_{S})} \prod_{j \neq a} \frac{\big( z_a e^{-i \theta_{j}} - z_j \big)^{3}}{\big( z_a - z_j \big)^{3}}.
\end{align}
Therefore, $\rho_j$ can be expressed as
\begin{align} 
  \frac{1}{\int \prod_{i} d^2 r_i  \big| 
  \Psi_{\frac{1}{3}} \big|^2} \int \prod_{i} d^2 r_i | \Psi_{\frac{1}{3}} |^2  \sum_{a=1}^{N_{S}} 
  \mathcal{Z}_{a} (\theta_{j}; \{ \vcr_{i} \}), 
\end{align}
where we have symmetrized $\mathcal{Z}$ over all particles to increase the rate of convergence. The above
expression has the same form as Eq.~(\ref{eq:Coulomb}) and can therefore be evaluated through very similar
Metropolis sampling.

\begin{figure}[t]
  \centering
  \includegraphics[width=\columnwidth]{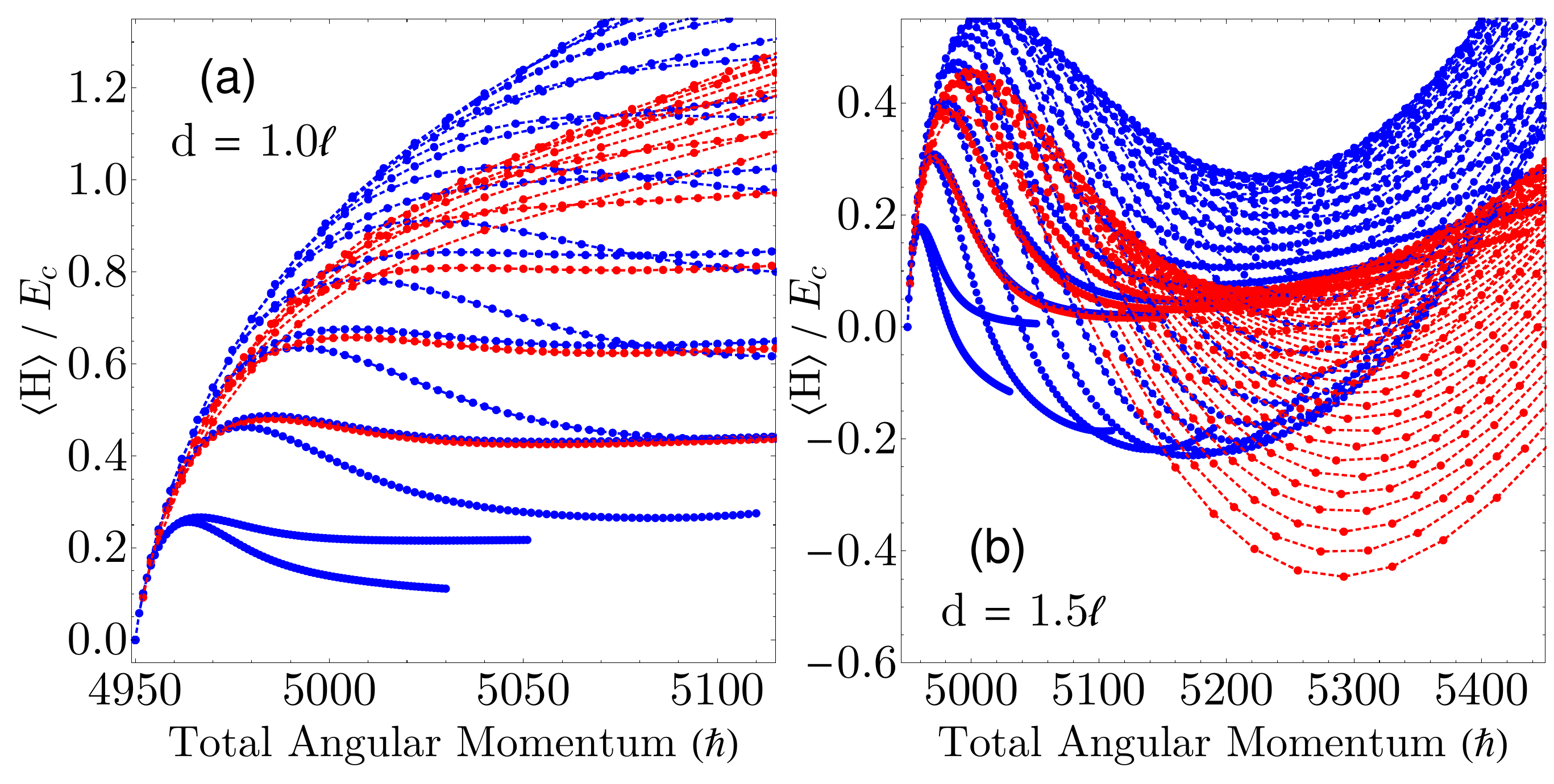}
  \caption{ Results of the variational analysis using $100$ electrons with 
  a charge-neutral confining potential. The blue (red) dots show the total energy 
  of the variational states with a $\nu = 1$ integer ($\nu = \frac{1}{3}$ fractional) 
  side-strip as a function of the total angular momentum for a (a) sharp ($d = \ell$) 
  and (b) smooth ($d = 1.5 \ell$) confining potential. 
  Each curve corresponds to states with the same separation between the bulk and side-strip ($L_{S}$)
  but with different number of electrons in the side-strip ($N_{S}$). The curves shown here correspond
  to $L_{S}$ varying from $0$ to $30$ guiding centers.
  The energy of the unreconstructed state has been subtracted to make comparison easier.
  (a) For sharp edges $(d < 1.3 \ell)$ the ground state is the one with minimum angular momentum, 
  implying no edge reconstruction. (b) For smooth edges $(d > 1.3 \ell)$ the ground state shifts to 
  a higher angular momentum sector, implying that the electronic disk expands and the edge 
  undergoes reconstruction. The minimum energy state lies on the curve corresponding to $L_{S} = 0$. 
  Panel (b) shows that a fractional reconstruction is energetically favorable to
  an integer reconstruction. }  
\end{figure}

\section{III.\,\,\,\,\,\,      Charge Neutral Confining Potential}

In the main text, the edge confining potential is modelled as a ramp function which interpolates linearly 
between two constants. Since any fairly smooth edge potential can be linearized around the chemical potential, 
we do not expect the results of our variational analysis to be modified by using a different smooth potential.

In order to verify this claim and properly compare our results with existing literature, we have repeated our 
variational analysis with a commonly used charge-neutral edge potential~\cite{SKunYangIQHS,SKunYang_2002,SKunYang_2003,
SKunYang_2008,SKunYang_2009}. 
Specifically, the confining potential is modelled as the electrostatic potential of a positively charged background 
disk separated by a distance $d$ from the electron gas along the direction of the magnetic field. Therefore 
the confining potential defined in Eq.~(1) of the main text is replaced by
\begin{align}
  V_{c} (r) = \int_{0}^{R} dr^{\prime} \int_{0}^{2\pi} d \theta \frac{E_{c} \sigma}{\sqrt{d^2 + r^2 + {r^{\prime}}^2 - 
  2 r^{\prime} r \cos \theta}}
\end{align}
where, the density ($\sigma$) and the radius ($R$) of the background disk depend on the bulk filling factor $(\nu_B)$ and 
number of electrons ($N_{S} + N_{S}$) respectively. Charge neutrality of the full system requires $\sigma = \nu_B / 2\pi \ell^2$ and 
$R^2 = 2(N_{S} + N_{B}) \ell^2$ (where $\ell$ is the magnetic length).
The resulting edge potential is quite sharp at $d = 0$, and becomes smoother as $d$ increases. 

Fig.~S1 shows the total energies for the two class of variational states with a total of 100 electrons when this 
charge neutral potential is employed. The blue (red) dots correspond to different states with an integer 
(fractional) side-strip at the edge. The curves correspond to states with the same separation between the bulk and 
side-strip ($L_{S}$) and different number of electrons in the side-strip ($N_{S}$). Clearly for a sharp confining potential 
[$d < 1.3 \ell$, Fig.~S1(a)] the lowest energy state is the one with the minimal angular momentum (in this case $4950 \hbar$), 
which corresponds to the unreconstructed $\nu = 1$ state.

For smoother potentials [$d > 1.3 \ell$, Fig.~S1(b)] the lowest energy state has a much larger angular momentum
($5292 \hbar$ for $d = 1.5 \ell$ with $N_{S} = 19$ and $L_S = 0$) than the compact state implying that the edge 
has undergone reconstruction. The states with a fractional edge are found to have a lower
energy than the states with an integer edge for sufficiently smooth potentials. Thus the central result of this work
is unaffected by the specific choice of confining potential.

\begin{figure}[t]
  \centering
  \includegraphics[width=\columnwidth]{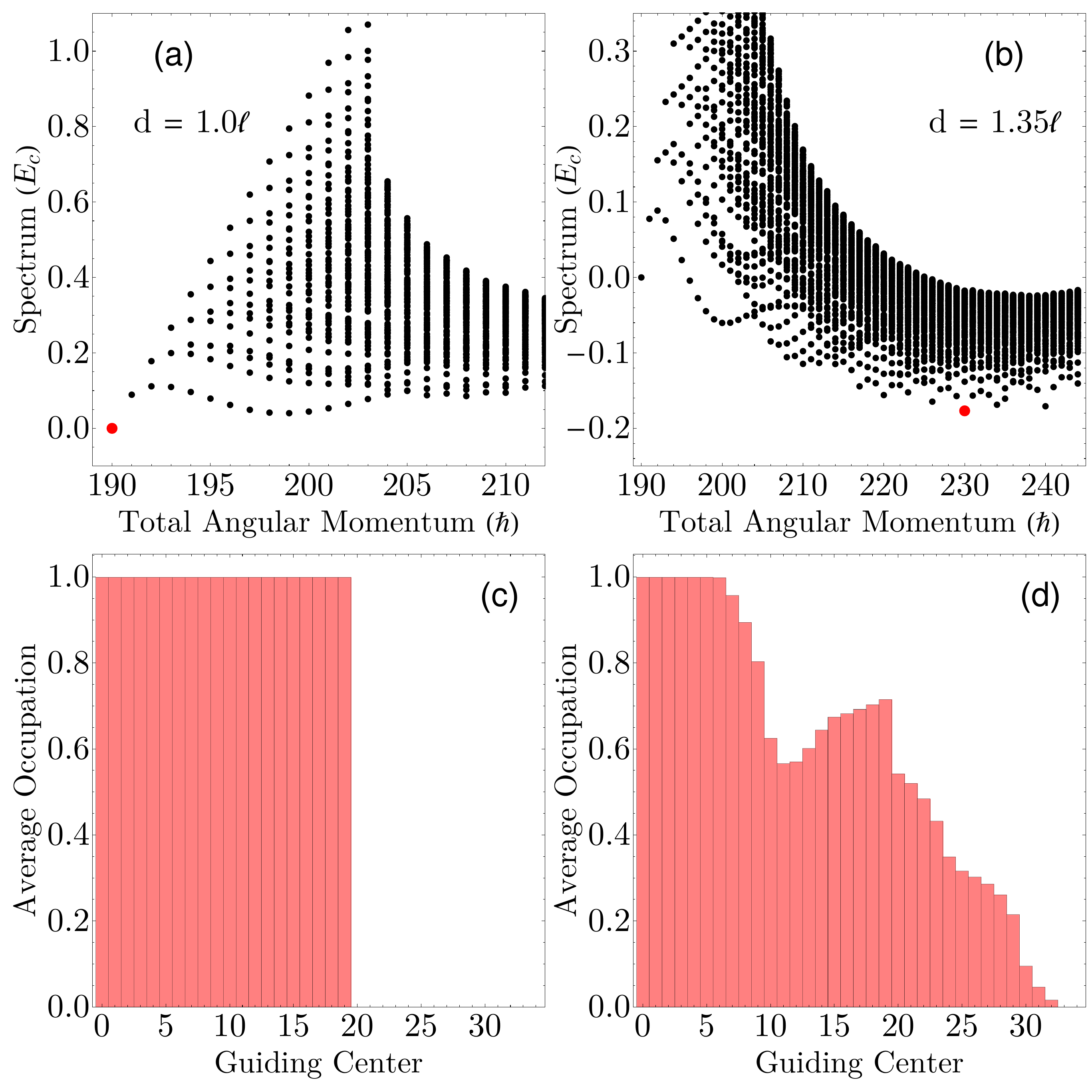}
  \caption{ Results from exact diagonalization of the $\nu = 1$ phase with 20 electrons restricted to 
  35 guiding centers.
  (a-b) The black dots shows the low energy spectrum as a function of
  the total angular momentum for a (a) sharp ($d = 1.0 \ell$) and (b) smooth ($d = 1.35 \ell$)
  confining potential (here $\ell$ is the magnetic length). The red dot corresponds to the exact ground state.
  The energy of the unreconstructed state has been subtracted
  to make comparison easier. (a) For sharp confining potentials, the ground state is in the lowest 
  allowed angular momentum sector, implying no edge reconstruction. (b) For sufficiently smooth 
  confining potentials, the ground state shifts to a larger angular momentum sector, implying the edge has undergone reconstruction.
  (c-d) depict the average occupation of the guiding centers in the exact ground state 
  at (c) $d = 1.0 \ell$ and (d) $d = 1.35 \ell$. (d) shows that the filling factor at the reconstructed
  edge (and even its precise location) cannot be concluded from this analysis. } 
\end{figure}

\section{IV.\,\,\,\,\,\,      Exact Diagonalization Analysis}

In this section, we present an exact diagonalization (ED) analysis of the edge of $\nu = 1$ phase. 
As described in the main text, we consider spinless electrons in the disk geometry and neglect higher 
Landau levels. Then the Hamiltonian is composed of a circularly symmetric one-body confining potential 
and the two-body Coulomb repulsion. Here we employ the charge-neutral confining potential described in 
Section III~\cite{SKunYangIQHS,SKunYang_2002,SKunYang_2003,
SKunYang_2008,SKunYang_2009}. We include up to 20 electrons in 35 guiding centers ($m = 0$ to $34$) and 
perform ED to find the low energy spectrum in several angular momentum sectors. Note that a relatively large number 
of electrons (as far as ED is concerned) is possible in our case because the bulk filling factor is 1. The 
average occupation of each guiding center is readily found from the ground state wavefunction. 

Fig.~S2 shows the spectrum and ground state occupations for $20$ electrons at two different confining potentials. 
In Figs.~S2(a) and (b) the black dots show the low energy spectrum as a function of the total angular momentum 
while the red dot corresponds to the exact ground state. For a sharp confining potential [Fig.~S2(a)] the ground
state is the one with minimal angular momentum ($190 \hbar$ for $20$ electrons). Fig.~S2(c) shows that the average occupation
of guiding centers in the ground state drops sharply from $1$ to $0$ at the edge. Clearly, this corresponds to the unreconstructed 
$\nu = 1$ state. 

For smoother confining potentials ($d > 1.1 \ell$, where $\ell$ is the magnetic length) the ground state shifts 
to a higher angular momentum sector indicating that the edge has undergone reconstruction. Fig.~S2(b) shows 
that for $d = 1.35 \ell$, the ground state has angular momentum $230 \hbar$ for which the average 
occupation of guiding centers [Fig.~S2(d)] falls smoothly and non-monotonically from $1$ to $0$ at the edge. 
However, while the occupation of guiding centers close to the origin is $\sim 1$, the filling factor of the edge 
is far from quantized. Even the precise location of the edge is blurred. Both these issues arise 
due to the small number of electrons included in this analysis.

As is evident from this discussion, even 20 electrons are insufficient to make any conclusion regarding the precise
filling factor of the side-strip formed at the edge during reconstruction or the nature of the emergent counter-propagating
edge modes. This is a serious limitation of the ED method. Our variational analysis, on the other hand, allows us to address
the problem of edge reconstruction within a quantum many-body framework using a very large number of electrons
($100$ in the current manuscript), and our results clearly indicate that fractional edge reconstruction is 
energetically favorable compared to integer reconstruction.


\begin{thebibliography}{49}
\bibitem [{fno()}]{fnoteLs}%
  \BibitemOpen
  \href@noop {} {}\bibinfo {note} {We find that $L_{S} \leq 2$ throughout the
  parameter space. Consequently, the electronic density shows that the two
  droplets are smoothly connected for all values of the parameters [as in
  Fig.~2(d)]. This is consistent with the exact diagonalization results of
  Chamon and Wen~[5].}\BibitemShut {Stop}%
\bibitem [{Not()}]{Note2by3}%
  \BibitemOpen
  \href@noop {} {}\bibinfo {note} {We also performed a variational calculation
  for an edge structure where the filling factors follow the profile (from bulk
  to edge) $1 \rightarrow 2/3 \rightarrow 1 \rightarrow 0$, such that the edge
  has a similar structure to the $\nu = 2/3$ state. However, that configuration
  turned out to be energetically unfavorable compared with the structure in
  Fig.~1(b).}\BibitemShut {Stop}%
\bibitem [{The()}]{ThermalTransport}%
  \BibitemOpen
  \href@noop {} {}\bibinfo {note} {The thermal Hall conductance is $3 \times
  \pi^2 k_B^2 T/3 h$ for an unequilibriated edge for both integer and
  fractional reconstructions since it only depends on the number of chirals
  participating in transport. For an equilibriated edge it reduces to $\pi^2
  k_B^2 T/3 h$, as expected for the $\nu = 1$ QH state.}\BibitemShut {Stop}%
\end{thebibliography}
\end{document}